\documentclass[10pt]{article}
\usepackage[cp1251]{inputenc}
\usepackage{cite}
\usepackage[dvips]{graphicx}
\usepackage[dvips]{epsfig}   \usepackage[dvips]{changebar}

\usepackage{caption2}

\begin{document}
\title{Possible nature of the electric activity of
   He II\\ observed in experiments with second sound}
\author{V. M. Loktev, M. D. Tomchenko
\bigskip \\ {\small
   Bogoliubov Institute for Theoretical Physics, NAS of Ukraine,} \\
  {\small 14-b, Metrologichna str., Kiev, 03143, Ukraine} \\
    {\small E-mail:vloktev@bitp.kiev.ua; mtomchenko@bitp.kiev.ua}}
   \date{\empty}
 \maketitle
 \large
 \sloppy

  \textit{An attempt is made to explain the nature of the electric signal
    observed in He II in a second-sound standing wave.
    Using the general quantum-mechanical
    principles, we show that, due to interatomic interaction,
    each atom of He-II acquires a small fluctuating induced dipole moment
    with the average value of its modulus $\bar{d}_{at} =
   2e\delta$,  $\delta \simeq 2.6 \times 10^{-4} \mbox{\AA}$.
   A directed flux of microscopic vortex rings --- which, together with phonons,
   are thermal excitations of He II --- forms in the
   second-sound standing halfwave.
    This flux partially orders the chaotically oriented dipole moments
   of the atoms, which results in volume polarization of He II. The observed
   electric induction $\triangle U \approx k_B \triangle T/2e$
   can be explained theoretically under the assumption that each vortex
   ring possesses a dipole moment $d_{vr}$ of the order of ten atomic
   moments, $d_{vr} \sim 10 \bar{d}_{at}$.}

   \textit{It is shown also that the theoretical  value of the voltage  $\triangle U$ induced in
   He II by the volume system of dipoles strongly depends on the dimensions of the resonator
   for dipoles of any origin. The experimental value of $\triangle U$
   is the same for two resonators of different size; therefore, this voltage may be not connected
     with the volume polarization of He II.}\\
   PACS: 67.40.Pm \\
     \textit{Keywords}: Helium-II, electric activity, dipole moment,
  vortex rings. \\

   \section{Introduction}

    A series of experiments were recently carried out
    \cite{rub,rub1,rub2}, in which electric properties of He II have
    been discovered. This is rather unexpected, given that He II is a
    dielectric and that a free He$^4$ atom does not possess either dipole
    or higher multipole moments. Several models have been proposed  to explain
    the electric activity of He II \cite{mel,nac,push}, but
    each of them encounters with certain difficulties and does not explain
    the effects of \cite{rub,rub1,rub2} in a plausible way.  In
    this work, we study a possible nature of the electric activity of
   He II observed by Rybalko \cite{rub} in experiments with second sound.
   More details of this analysis are presented in \cite{lt}.

   \section{Induced dipole moment of He II atoms}

   It was found in the experiment \cite{rub} that the second-sound
   standing half-wave induces an alternating voltage $U$ in He II
   with the frequency $\omega_2$ of the second sound. It   was
   established that, in the temperature range $T=1.4$--$2\,$K, the
   amplitudes of the voltage ($\triangle U$) and the temperature ($\triangle T$)
   are related as $\triangle
U/\triangle T \approx k_b /2e$.

We assume that the effect observed in \cite{rub} is connected with volume polarization of
helium-II, which is caused by the existence of microscopic dipole moments (DM).  Its
origin should be clarified. It is shown in \cite{mel} that atoms of He II possess an
inertial DM caused by their acceleration in sound waves. However, such DM give
polarization which is two to three orders of magnitude smaller than the observed value.
Below we show that helium atoms possess not only inertial DM but also induced DM caused
by interaction between neighboring atoms.

According to the standard perturbation theory \cite{lt}, the ground-state wave function
of two interacting helium-II atoms $A$ and $B$ has the form
          \begin{equation}
  \Psi^{AB}_0 = c_0 \Psi^{A}_0\Psi^{B}_0 + c_1
  \Psi^{A}_1\Psi^{B}_0 + c_1 \Psi^{A}_0\Psi^{B}_1 + c_2
  \Psi^{A}_1\Psi^{B}_1 + \ldots,
    \label{psi0-2}     \end{equation}
where $\Psi_0$ is the $1s^2$ ground-state wave function of the He$^4$ atom, and $\Psi_1$
is the excited state $1s2p$ with  $l=1$ and $m=0$.  We neglect the next corrections in
(\ref{psi0-2}). In the quadrupole approximation for the perturbing potential, in the
first order of perturbation theory, we obtain \cite{lt}
     $c_1=-8.15 a_{B}a^3/R^4$,  $c_2 =
     1.43 a_{B}a^2/R^3$, where $R$ is the distance between atoms,
     $a=a_B/Z^* =0.313 \mbox{\AA}$, $Z^* = 2-5/16$, and
     $a_B = \hbar^2/me^2=0.529 \mbox{\AA}$ is the Bohr
     radius. We calculate the DM of atom $A$ according to the
     general formula
        \begin{equation}
  \textbf{\emph{d}}_0^A = \int \Psi^{*AB}_0 e(\textbf{\emph{r}}_1^A + \textbf{\emph{r}}_2^A)
   \Psi^{AB}_0 d\textbf{\emph{r}}_1^A d\textbf{\emph{r}}_2^A d\textbf{\emph{r}}_1^B
   d\textbf{\emph{r}}_2^B,
    \label{dip0}     \end{equation}
    where $e$ is the electron charge. From (\ref{psi0-2}) and (\ref{dip0}),
     for $R$ equal to the average distance between the He II atoms ($3.6\,\mbox{\AA}$), we obtain:
      \begin{equation}
  \textbf{\emph{d}}_0    \approx
   -2e\delta\frac{\textbf{\emph{R}}}{R},  \qquad \delta = 2.63 \times 10^{-4} \mbox{\AA}.
    \label{dipk}     \end{equation}
The DM arises as a result of stretching of the electronic cloud of an atom in the
direction away from the neighboring atom.

Due to interaction with neighbors and nonuniform distribution of atoms, each atom of He
II acquires certain DM $\textbf{\emph{d}}_{at}$. The direction and magnitude of
$\textbf{\emph{d}}_{at}$ are different for different atoms, but the average value
        is $\bar{d}_{at} \sim d_{0}$.

   \section{Focusing of the induced atomic dipole moments by vortex rings}

In a second-sound standing half-wave, the temperature varies according to the law
       \begin{equation}
  T = T_0 - 0.5\triangle T \cos(\omega_2 t)\cos(z\pi/L),
    \label{T}     \end{equation}
where $z$ is the coordinate along the resonator, and  $L$ is the resonator length. By the
reasoning of \cite{lt}, microscopic vortex rings, being a type of He II thermal
excitations \cite{ring}, can possess their own DM $d_{vr}$, with values which are thus
far unknown. An ordered flux of vortex rings (along the direction of decreasing $T$) is
present in a second-sound standing half-wave due to the temperature difference. Because
vortex rings possess DM, this flux creates the polarization of helium
        \begin{equation}
  P_z(z) =  \frac{\eta}{L}\cos(\omega_2 t)\sin{\left (\frac{z\pi}{L}\right )}, \qquad
  \eta= \frac{\pi}{16\varepsilon_{He}}\frac{\partial n_{vr}(T)}{\partial T}
              d_{vr} n^{-1/3}_{vr} \triangle T.
      \label{pl22}     \end{equation}
Here, $ n_{vr}(T)$ is the equilibrium distribution of rings \cite{ring}. Such
polarization induces the following voltage $U$ between the end walls of the resonator:
       \begin{equation}
  U =  \eta\cos(\omega_2 t)  \gamma (R,L),
      \label{du}     \end{equation}
where $\gamma (R,L)$ is the factor taking into account the boundary conditions. For a
short resonator in \cite{rub}, $\gamma \approx 1.38$; for a long one, $\gamma \approx
1/20$; for the infinite medium, we have $\gamma=4$. It follows from  (\ref{du}) that, at
$T=1.4\,$K,
     \begin{equation}
\frac{\triangle U}{\triangle T} \approx \frac{k_B}{2e}\,
  \frac{ \gamma (R,L)d_{vr}}{3,9 \bar{d}_{at}}.
  \label{dU3}  \end{equation}
The experimental value $\small{\triangle} U/\triangle T \approx k_B/2e$ is obtained from
(\ref{dU3}) at $d_{vr} \sim 10 \bar{d}_{at}$.  However, the value of $\small{\triangle}
U/\triangle T $ strongly depends on the boundary conditions, which should be checked
experimentally.  The indicated dependence of $\gamma (R,L)$  on the boundary conditions
takes place for DM of any origin, causing polarization of helium.  The absence of this
dependence in the experiment will indicate that the observed value of $\small{\triangle}
U$ either is connected with certain boundary effect in He II or is not connect with He II
at all, but results from one kind of thermo-e.m.f.

     \section{Summary}

We proposed the idea that the electric signal $\small{\triangle} U $ observed in a
standing wave of second sound is caused by a directed flux of vortex rings possessing
proper dipole moment $ \sim 10 \bar{d}_{at} \sim e\,5\times 10^{-3} \mbox{\AA}$. A ring
may possess DM due to asymmetry in the location of atoms in front of and behind the ring.
It is shown that, in the case of the volume nature of $\small{\triangle} U $, the
theoretical value of $\small{\triangle}U$ should strongly depend on the dimensions of the
resonator, which effect can be tested experimentally.  Such a dependence was not observed
in \cite{rub}; therefore, the effect of \cite{rub} may be connected not with volume but
rather with the boundary properties of He II or with the electric properties of the
materials used to measure $\triangle U$.

    \section{Acknowledgements}

The authors are grateful to A.\,S.\,Rybalko and E.\,Ya.\,Rudavski\v{i} for the
explanation of the specific features of the experiment \cite{rub} and helpful remarks,
and to Yu.\,V.\,Shtanov for valuable discussion.

\renewcommand\refname{REFERENCES}

       \end{document}